\documentclass[PRA]{article}

\usepackage[utf8]{inputenc}

\usepackage{latexsym}
\usepackage{verbatim}
\usepackage[a4paper, total={6.5in, 9in}]{geometry}
\usepackage{silence}
\usepackage{amsthm,amsmath,amsfonts,amssymb,verbatim, color}
\usepackage{graphicx}
\usepackage[percent]{overpic}
\usepackage{bm}
\usepackage{braket}
\usepackage{float}
\usepackage{mathtools}
\usepackage{appendix}
\usepackage[colorlinks=true,citecolor=blue,linkcolor=black,urlcolor=blue]{hyperref}
\usepackage{graphics} 
\graphicspath{ {figures/} }
\usepackage[caption=false]{subfig}

\usepackage[mathscr]{euscript}
\usepackage[export]{adjustbox}
\usepackage[bottom]{footmisc}
\usepackage{textcomp}
\usepackage{braket}

\begin{document}

\title{Controlling and engineering a quantum state in a multi-qubit system employing the quantum Zeno effect}
\author{$^{1}$Dhruva Naik, $^{2}$Garima Rajpoot, and $^{3}$Sudhir Ranjan Jain \\ 
$^{1}${\small{\it Department of Physics, IIT Guwahati}}\\{\small{\it Amingaon, Guwahati, 781039, India}}\\
$^{2}${\small{\it Theoretical Nuclear Physics and Quantum Computing Section}}\\ {\small{\it Nuclear Physics Division, Bhabha Atomic Research Centre, Mumbai 400085, India}}\\
$^{3}${\small{\it UM-DAE Centre for Excellence in Basic Sciences, University of Mumbai}}\\ {\small{\it Vidyanagari campus, Mumbai 400098, India}}\\
}
\date{}
\maketitle
\begin{abstract}
    Controlling quantum jumps is crucial for reliable quantum computing. In this work, we demonstrate how the quantum Zeno effect can be applied to a two-qubit system interacting with an ancilla—a component of surface code architecture—to control undesired transitions. Further, we show that by designing the interaction and tuning measurement frequency, the quantum Zeno effect can be used to achieve the desired target state in both single-qubit and two-qubit systems.
\end{abstract}
\section{Introduction}
The state of a qubit can be represented on a Bloch sphere where the state's transition is continuous. For fault-tolerant computation, monitoring and controlling these transitions is important \cite{Dorit}. The associate quantum jumps were observed in a single $Ba^+$ ion using LASER \cite{Dehmelt}. Based on Dehmelt's shelving scheme \cite{Dehmelt, Dehmelt2}, an experiment on catching and reversing quantum jump has been demonstrated successfully by Minev et al.\cite{Minev}. Monitoring of the qubit dynamics can be done either by projective measurement or weak measurement. Projective measurement is an instantaneous ideal measurement where the response time of the measuring apparatus is shorter than other relevant time scales \cite{Koshino, braginsky} and the system collapses to an eigenstate of the measurement operator. For reliable computation, monitoring of a qubit state without disturbing its state is required. Thus we perform a weak measurement - an instantaneous measurement of the detector (probe) that is weakly interacting with the qubit \cite{braginsky}. The dynamics of a qubit system interacting with a probe field has been approached using the theory of continuous measurement to get the stochastic master equation by discretizing the interaction in time \cite{Gross}. The solution of this master equation consists of a multitude of paths with different weights that constitute the quantum trajectories \cite{Carmichael}. Due to entanglement, the measurement of the probe affects the evolution of the system.  To accommodate the measurement back action, stochastic path integral formalism (CDJ formalism) has been developed by using the joint probability density function of measurement outcomes to study the dynamics of a qubit under weak, continuous, repeated measurements \cite{CDJ}. This formalism includes the extremization of action which gives the most likely paths with boundary conditions defined by preselected \& postselected states as solutions to a set of ordinary differential equations. Using this approach the stochastic evolution equations for position and momentum expectation values for a simple harmonic oscillator have been deduced in \cite{KLJ}. The mathematical similarity between the action formalism of most likely quantum path and ray optics has been drawn in an experiment \cite{Naghiloo} where homodyne detection of resonance fluorescence from a superconducting artificial atom was performed by employing phase-sensitive amplification. It was observed that multiple most likely paths in experimental data were in agreement with the theoretical prediction.

The continuously repeated projective, instantaneous measurements on a system lead to freezing of state, known as the quantum Zeno effect (QZE)\cite{SM, Peres}. To understand the QZE, the path-integral representation of the reduced density operator is presented in \cite{Onofrio1, Onofrio2, Halliwell}. The QZE helps control the state of the qubit. It has been used to deterministically generate different multiparticle entangled states in an ensemble of $36$ qubit atoms in less than $5\mu$s \cite{Giovanni}. A detailed study of the dynamics of a quantum system in the quantum Zeno regime has been done for a qubit \cite{krj} and a qutrit \cite{Komal}. 

In \cite{krj}, it has been shown that if the detection frequency is smaller than the Rabi frequency of the qubit, the oscillations slow down, eventually coming to a halt at a resonance of interest when measurements are spaced exactly by the time of transition between the two states. The enhancement in transition time due to the measurement provides time to detect and correct the error. On the other hand, in the limit of a large number of repeated measurements, a cascade of stages occurs \cite{Parveen} and the transition points are saddle points in the phase space representation. The pair of expanding and contracting directions in phase space are inverted for the two points \cite{krj}. There also appears a pair of separatrices. Altogether, the phase space is divided by these special structures, guiding the phase flow and the corresponding transitions and jumps in a rather intricate and instructive way. In the quantum Zeno regime, the qubit state is localized near one saddle point and delocalized near the other saddle point resulting in the freezing of state near the stable transition point. 

In Section \ref{Two_qubit}, we present a way to shelve the state of a two-qubit system interacting with a single ancilla using weak measurement. It is a step towards understanding the quantum Zeno effect in the large structure of quantum error-correcting surface codes \cite{Fowler, Genus2, Genus}.  Weak measurement helps in engineering the target state \cite{Parveen2}. Using this approach, in Section \ref{target state}, we have shown that the quantum Zeno effect helps in achieving a desired target state by providing the conditions on the interaction between the qubit \& ancilla and the measurement of the ancilla. 

\section{System Dynamics}\label{Two_qubit}
Let us consider a system of two qubits, undergoing coherent oscillations with Rabi frequencies $2\omega_1$ and  $2\omega_2$. The state $\ket{1}$ of both systems are continuously being monitored by the detector (ancilla). The detector is subjected to measurements at successive time intervals, separated by $\delta t$. The detector is another two-level system, initially prepared in the eigenstate, $\ket{0}_d$ of $\sigma_{z_d}$ with eigenvalue $1$. We measure $\sigma_{y_d}$ of the ancilla, after which it is reset to $\ket{0}_d$. The total Hamiltonian of both the qubits with the interaction is $H=H_s+H_{int}$:
\begin{equation}
    H=\omega_1 \sigma_{x_1}\otimes {\mathbb{I}}_2\otimes {\mathbb{I}}_d + \omega_2 {\mathbb{I}}_1\otimes \sigma_{x_2}\otimes {\mathbb{I}}_d+\frac{J_1}{2} ({\mathbb{I}}_1-\sigma_{z_1})\otimes {\mathbb{I}}_2\otimes \sigma_{y_d} +\frac{J_2}{2}{\mathbb{I}}_1\otimes ({\mathbb{I}}_2-\sigma_{z_2})\otimes  \sigma_{y_d} 
\end{equation}
where the subscripts $1(2)$ and $d$ stand for qubit-$1$ (qubit-$2$) and detector respectively. $J_1 (J_2)$ is the coupling strength between the detector and the qubit-$1$ (qubit-$2$ ) which can be written as $J_1$=$\sqrt{\frac{\alpha_1}{\delta t}}$ ($J_2$=$\sqrt{\frac{\alpha_2}{\delta t}}$) \cite{Martin} and $\mathbb{I}$ is $2\times 2$ identity matrix. The possible measurement outcome, $r$, can be $0$ or $1$. Hence, the system is evolving under the combined effect of its Hamiltonian and the coupling to the detector which is given by the unitary evolution. For the detector state to be in $|r\rangle_d$ and in the scaling limit for continuous measurement, $\delta t\to 0$, $J^2\delta t\to\alpha=\rm{constant}$, the total density matrix of the two-qubit system is \cite{krj, Martin}
\begin{equation}
   \rho(t+\delta t)=\frac{M^r U \rho(t)U^\dagger M^{r^{\dagger}}}{Tr[M^r U \rho(t)U^\dagger M^{r^{\dagger}}]}
\end{equation}
where $U=e^{-\iota H\delta t}$ describes the unitary evolution of the system. $M^{(r)}$ is the Kraus operator \cite{Kraus}:
\begin{alignat}{1}
    M^r &={\mathbb I}_1\otimes {\mathbb I}_2 \otimes ~_{d}\bra{r}U_{s-d}\ket{i}_d{\mathbb I}_1\otimes {\mathbb I}_2,
\end{alignat}
where $\ket{i}_d$ is the initial state of the ancilla, $U=e^{-iH_s \delta t}$ and $U_{s-d}=e^{-iH_{int}\delta t}$. Considering the post-selected measurement for $r=0$:\begin{alignat}{1}\label{rho12t}
   \rho(t+\delta t)&=\frac{M^0 U \rho(t)U^\dagger M^{0^{\dagger}}}{Tr[M^0 U \rho(t)U^\dagger M^{0^{\dagger}}]},
\end{alignat}
where the denominator normalizes the density. After each measurement, we initialize the ancilla but the two qubits which are being monitored remain entangled. We write the general expression of the density matrix of a two-qubit system with an ancilla at time $t$ as \cite{Aryeh}:
\begin{alignat}{1}\label{rhot}
    \rho&(t)=\frac{1}{4}\bigg({\mathbb I}\otimes {\mathbb I}+\rho_1\otimes {\mathbb I}+{\mathbb I}\otimes \rho_2 + e_{11}\sigma_{x_1}\otimes\sigma_{x_2}+e_{12}\sigma_{x_1}\otimes\sigma_{y_2}+e_{13}\sigma_{x_1}\otimes\sigma_{z_2}+e_{21}\sigma_{y_1}\otimes\sigma_{x_2}\nonumber\\     &+e_{22}\sigma_{y_1}\otimes\sigma_{y_2}+e_{23}\sigma_{y_1}\otimes\sigma_{z_2}+e_{31}\sigma_{z_1}\otimes\sigma_{x_2}+e_{32}\sigma_{z_1}\otimes\sigma_{y_2}+e_{33}\sigma_{z_1}\otimes\sigma_{z_2}\bigg)\otimes\rho_d,
\end{alignat}
where $\rho_{1(2)}=(x_{1(2)}\sigma_{x_{1(2)}}+y_{1(2)}\sigma_{y_{1(2)}}+z_{1(2)}\sigma_{z_{1(2)}})$ with $x_{1(2)}, y_{1(2)}$, and $z_{1(2)}$ are the Bloch sphere coordinates of the two qubits and $e_{ij}(i,j=1,2,3)$ are the entanglement variables. Let us explain these entanglement variables by a simple example of Bell's state $\ket{\psi} = (1/\sqrt{2})(\ket{00} + \ket{11})$ which is a maximally entangled state. The density matrix of this state is:
\begin{alignat}{1}     
 \rho & = \frac{1}{2}\begin{pmatrix}
  1 & 0 & 0 & 1\\
  0 & 0 & 0 & 0\\
  0 & 0 & 0 & 0\\
  1 & 0 & 0 & 1
  \end{pmatrix}.
\end{alignat}
 The density matrix above cannot be written without considering the entanglement variables. The values of the Bloch sphere coordinates and entanglement variables for this Bell's state are $x_{1(2)} = y_{1(2)} = z_{1(2)} = 0$, $e_{11} = 1, e_{12} = e_{13} = e_{21} = e_{31} = 0, e_{22} = -1, e_{23} = e_{32} = 0$, and $e_{33} = 1$.
 
Eqs. \eqref{rho12t}, \eqref{rhot} can be employed to explicitly write the  update equations of the Bloch sphere coordinates and the entanglement variables as:
\begin{alignat}{1}
    \dot{x_1}(t)&=\frac{1}{2}\bigg[-x_1 z_1 \alpha_1 + \sqrt{\alpha_1 \alpha_2}(e_{13}-x_1 (z_1+z_2-e_{33}))+(-x_1 z_2+e_{13})\alpha_2 \bigg],\nonumber\\
    \dot{y_1}(t)&=\frac{1}{2}\bigg[e_{23} (\sqrt{\alpha_1 \alpha_2}+\alpha_2) +y_1(-\alpha_1 z_1-\sqrt{\alpha_1\alpha_2}(z_1+z_2-e_{33})-\alpha_2 z_2)-4\omega_1 z_1 \bigg],\nonumber\\
    \dot{z_1}(t)&=\frac{1}{2}\bigg[\alpha_1(1-z_1^2)-\sqrt{\alpha_1 \alpha_2}(1 + e_{33} (1+z_1) -z_1^2 - z_2 - z_1 z_2)-\alpha_2 (e_{33} - z_1 z_2) + 4\omega_1 y_1 \bigg],\nonumber\\
    \dot{x_2}(t)&=\frac{1}{2}\bigg[ \alpha_1 (-z_1x_2+e_{31}) + \sqrt{\alpha_1 \alpha_2}(e_{31}-x_2 (z_1+z_2-e_{33}))+\alpha_2(-x_2 z_2) \bigg],\nonumber\\   
   \dot{y_2}(t)&=\frac{1}{2}\bigg[(\sqrt{\alpha_1 \alpha_2}+\alpha_1)(e_{32} -z_1 y_2) + \sqrt{\alpha_1 \alpha_2} y_2 (-z_2+e_{33})-\alpha_2 y_2 z_2-4\omega_2 z_2 \bigg],\nonumber\\
   \dot{z_2}(t)&=\frac{1}{2}\bigg[\alpha_1 (-z_1 z_2 +e_{33})-\sqrt{\alpha_1 \alpha_2} (1+z_2)(-1+z_1 +z_2-e_{33})+\alpha_2-\alpha_2 z_2^2 +4\omega_2 y_2 \bigg],\nonumber\\   
   \dot{e}_{11}(t)&=\frac{1}{2}\bigg[-\alpha_1 e_{11}z_1 +\sqrt{\alpha_1\alpha_2}(e_{22}-e_{11}(z_1+z_2-e_{33}))-\alpha_2 z_2 e_{11}\bigg],\nonumber\\
   \dot{e}_{12}(t)&=\frac{1}{2}\bigg[-\alpha_1 e_{12}z_1 -\sqrt{\alpha_1\alpha_2}(e_{21}+e_{12}(z_1+z_2-e_{33}))-\alpha_2 z_2 e_{12}-4\omega_2 e_{13}\bigg],\nonumber\\
   \dot{e}_{13}(t)&=\frac{1}{2}\bigg[-\alpha_1 e_{13}z_1 +\sqrt{\alpha_1\alpha_2}(x_1-e_{13}(z_1+z_2-e_{33}))+\alpha_2 (x_1-z_2 e_{13})+4\omega_2 e_{12}\bigg],\nonumber\\    
   \dot{e}_{21}(t)&=\frac{1}{2}\bigg[-\alpha_1 e_{21}z_1 -\sqrt{\alpha_1\alpha_2}(e_{12}+e_{21}(z_1+z_2-e_{33}))-\alpha_2 z_2 e_{21}-4\omega_1 e_{31}\bigg],\nonumber\\
   \dot{e}_{22}(t)&=\frac{1}{2}\bigg[-\alpha_1 e_{22}z_1 +\sqrt{\alpha_1\alpha_2}(e_{11}-e_{22}(z_1+z_2-e_{33}))-\alpha_2 z_2 e_{22}-4(\omega_1 e_{32}+\omega_2 e_{23})\bigg],\nonumber
\end{alignat}
\begin{alignat}{1}\label{update}
    \dot{e}_{23}(t)&=\frac{1}{2}\bigg[-(\alpha_1+\sqrt{\alpha_1\alpha_2})z_1 e_{23}+\sqrt{\alpha_1\alpha_2} (-z_2+e_{33})e_{23}-\alpha_2 z_2 e_{23}+(\alpha_2+\sqrt{\alpha_1\alpha_2})y_1\nonumber\\
   &+4(-\omega_1 e_{33}+\omega_2 e_{22}) \bigg],\nonumber\\
   \dot{e}_{31}(t)&=\frac{1}{2}\bigg[\alpha_1 (x_2-e_{31}z_1) +\sqrt{\alpha_1\alpha_2}(x_2-e_{31}(z_1+z_2-e_{33}))-\alpha_2 z_2 e_{31}+4\omega_1 e_{21}\bigg],\nonumber\\
   \dot{e}_{32}(t)&=\frac{1}{2}\bigg[(\alpha_1+\sqrt{\alpha_1\alpha_2})y_2 +
   e_{32}\{-\alpha_1z_1-\sqrt{\alpha_1 \alpha_2}(z_1+z_2-e_{33})-\alpha_2 z_2\}+4(\omega_1 e_{22}-\omega_2 e_{33}) \bigg],\nonumber\\
   \dot{e}_{33}(t)&=\frac{1}{2}\bigg[\alpha_1 (z_2-z_1 e_{33})-\sqrt{\alpha_1 \alpha_2}(e_{33}-1)(-1+z_1+z_2-e_{33})+(z_1-z_2 e_{33})\alpha_2 \nonumber\\
   &+4(\omega_1 e_{23}+\omega_2 e_{32}) \bigg].
\end{alignat}
To find the most-likely path from the quantum trajectories, we use the CDJ formalism \cite{CDJ} which explains that if the state of the system before the $k th$ measurement is $\textbf{q}(t_{k})$, then the conditional probability distribution $P(r_k|\textbf{q}(t_{k}))$ of the measurement outcome $r_k$: $P(r_k|\textbf{q}(t_{k}))\propto\exp\{\delta t\mathcal{F}[\textbf{q}(t_{k}),r_k]+\mathcal{O}(\delta t^2)\}$. Here,  $\mathcal{F}[\textbf{q},r]$  is the functional which can be written as $\mathcal{F}[\textbf{q}(t_{k}),r_k]=\ln{P(r_k|\textbf{q}(t_{k}))}$. It is the coefficient of $\delta t$ in the expression of $Tr[M^0 U \rho(t)U^\dagger M^{0^{\dagger}}]$. For the system defined above, this comes out to be:
\begin{equation}
    \mathcal{F}[\textbf{q},r]=\frac{1}{2}\bigg[\alpha_1 (z_1-1)+\sqrt{\alpha_1\alpha_2}(-1+z_1+z_2-e_{33})+\alpha_2 (z_2-1)\bigg].
\end{equation}
From the updated equations of the generalized coordinates of the system (Eq. \ref{update}) and the expression of the functional, we write the Hamiltonian which includes the effect of measurement on the two-qubit system as:
\begin{alignat}{1}\label{eq:ham}
    \mathcal{H}(\textbf{p},\textbf{q},r)&=\textbf{p}\dot{\textbf{q}}+\mathcal{F}[\textbf{q},r]\nonumber\\    &=p_{x1}\dot{x_1}+p_{y1}\dot{y_1}+p_{z1}\dot{z_1}+p_{x2}\dot{x_2}+p_{y2}\dot{y_2}+p_{z2}\dot{z_2}+\sum_{i,j=1}^{3}p_{eij}\dot e_{ij}\nonumber\\
    &+\frac{1}{2}\bigg[\alpha_1 (z_1-1)+\sqrt{\alpha_1\alpha_2}(-1+z_1+z_2-e_{33})+\alpha_2 (z_2-1)\bigg].
\end{alignat} 
We can find the update equations of the generalized momenta using Hamilton's equation (see Appendix). 
\begin{figure}[!ht]
    \centering
    \subfloat[]{\includegraphics[width=0.48\textwidth]{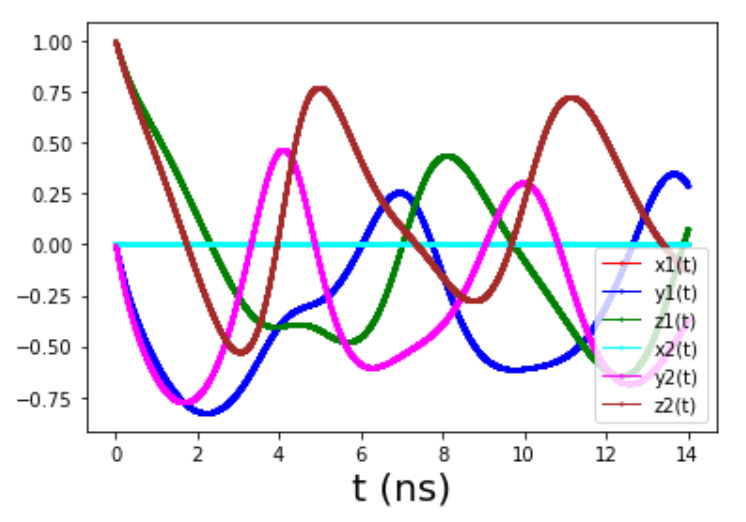}} 
    \subfloat[]{\includegraphics[width=0.483\textwidth]{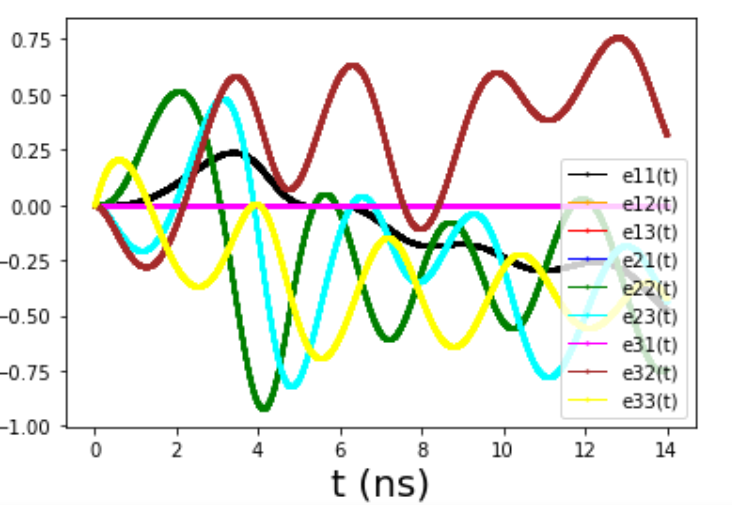}}\\
    \subfloat[]{\includegraphics[width=0.52\textwidth]{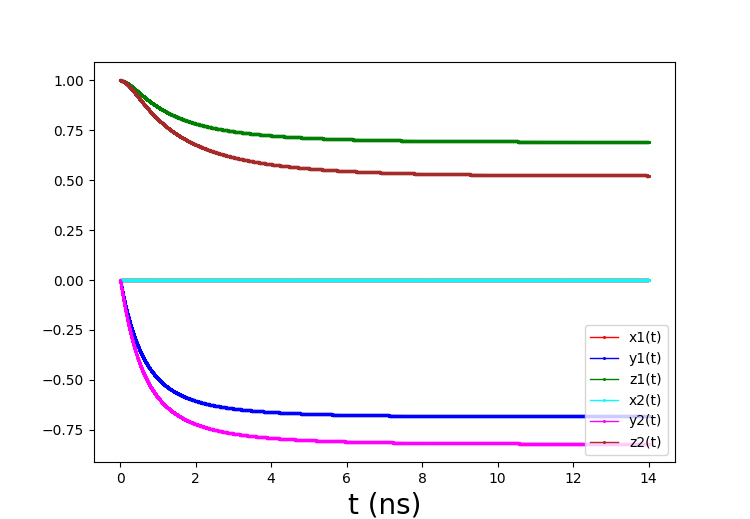}}
    \subfloat[]{\includegraphics[width=0.48\textwidth]{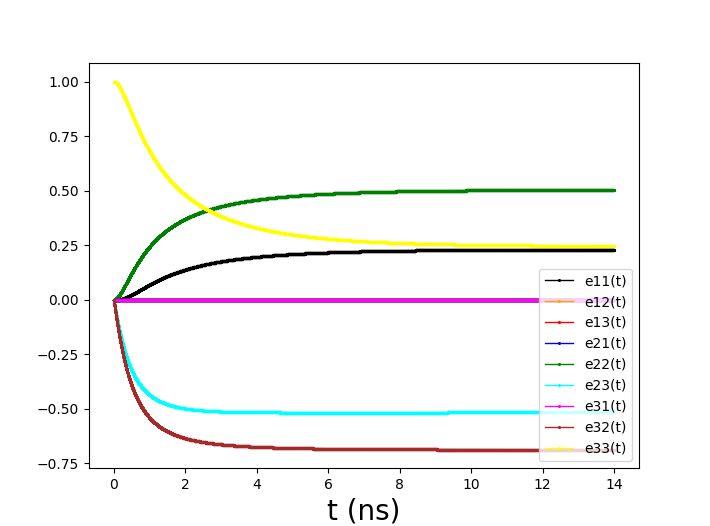}}
    \caption{Figure shows the dynamics for two qubits starting in a separable state, with initial conditions $x_1= x_2 = 0, y_1= y_2 = 0, z_1=\sqrt{1 - x_1^2 - y_1^2}, z_2 = \sqrt{1-x_2^2-y_2^2}$ and Rabi frequencies $2\omega_1 = 1, 2\omega_2 = 1.2$. The plots are shown for (a),(b) $\alpha_1=\alpha_2 = 0.5$ and in (c),(d) $\alpha_1=\alpha_2 = 3$. We can see that the Zeno regime sets in for higher values of $ \alpha$.}
    \label{fig:two_qubit}
\end{figure}
\begin{figure}[!ht]
    \centering
    \subfloat[]{\includegraphics[width=0.33\linewidth]{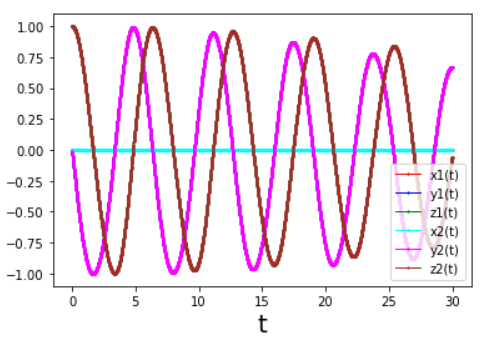}}
    \subfloat[]{\includegraphics[width=0.33\linewidth]{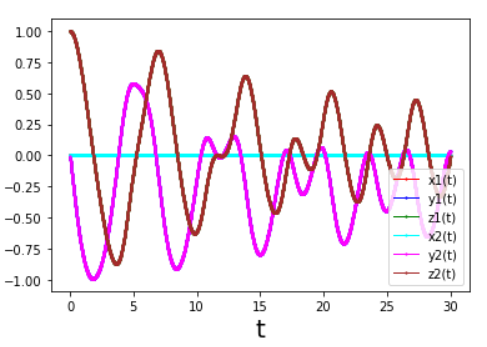}} 
    \subfloat[]{\includegraphics[width=0.33\linewidth]{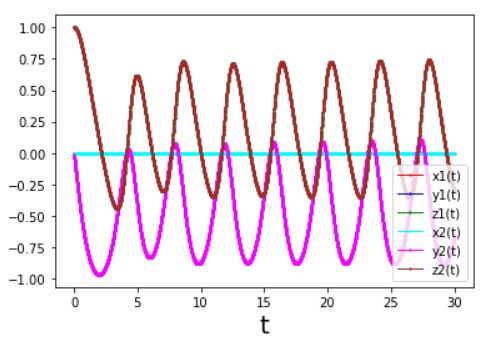}}  
    \caption{Figure shows the dynamics for the Rabi frequencies $2\omega_1=2\omega_2= 1$, with both the qubits starting in the $\ket{0}$ state. The plots for $\alpha_1=\alpha_2=$ (a) 0.1, (b) 0.3, and (c) 0.6.}
    \label{fig:same_freq}
\end{figure}
The dynamics of the position coordinates of the qubit with time are shown in Fig. \ref{fig:two_qubit}, \ref{fig:same_freq}. When the detection frequencies (corresponding to $\alpha_1$ and $\alpha_2$) are lower compared to the Rabi frequencies of the system ($\alpha_{1,2} < 2 \omega_{1,2}$ taken as $\alpha_{1,2} = 0.5$), we see that the system performs oscillations, with reduced frequencies (Fig. \ref{fig:two_qubit}(a,b), \ref{fig:same_freq}). So the time of transition is prolonged due to weak measurement. As shown in Fig. \ref{fig:same_freq}, an interesting phenomenon was observed at low measurement frequencies. Initially, the oscillation frequency decreases with increasing $\alpha$, but then the oscillations seem to split in two and the frequency doubles before it approaches the Zeno regime. 

When the detection frequency is higher ($\alpha_{1,2} = 3 > \omega_{1,2}$), we enter the Zeno regime and the coordinates stabilize at a certain value (Fig.\ref{fig:two_qubit}(c,d)). Nonetheless, quantum fluctuations and tunneling across the separatrices allow the system to keep evolving. The probability of tunneling is proportional to $\exp [-\mathrm \tau S_I/\hbar]$, where $S_I$ represents the imaginary part of the action as the separatrix is crossed, and $\tau$ is the inverse of the characteristic stability exponent along the unstable direction \cite{Alonso, Rakesh}, identified through the linearization process.

The Stochastic action integral for the phase space curves generated by the dynamics governed by the Hamiltonian (\ref{eq:ham}) is given by:
\begin{equation}\label{Action}
    \mathcal{S} (\textbf{p},\textbf{q},r)= \int_{t_i}^{t_f}(-\textbf{p}\dot{\textbf{q}}+\mathcal{H}(\textbf{p},\textbf{q},r))dt.
\end{equation}
Extremization of action (differentiating $\mathcal{S} (\textbf{p},\textbf{q},r)$ with respect to \textbf{p}, \textbf{q} and $r$, then equating it to zero to get $(\bar{\textbf{q}}, \bar{\textbf{p}}, \bar{r})$), gives the most optimal path. For the most optimal path, denoted by $(\bar{\textbf{q}}, \bar{\textbf{p}}, \bar{r})$, the Hamiltonian ${\mathcal H}(\bar{\textbf{q}}, \bar{\textbf{p}}, \bar{r})$ is a constant of the motion. Depending on the second variation of action, the most optimal path can be a local minimum, a local maximum, or a saddle point in the constrained probability space.
\subsection{Entanglement Entropy}
To study the entanglement between the two qubits as a function of  time, we trace over the generalized density matrix by the second qubit to obtain the reduced density matrix of the first qubit. This is employed to obtain the von Neumann entropy. If the eigenvalues of the reduced density matrix $\rho _1(t)$ are denoted by $\lambda_i(t)$ ($i=1,2$), then the von Neumann entropy \cite{David} is given by:
\begin{equation}
S(t) = -\sum_{i=1}^{2}|\lambda_i(t)|\hspace{1mm}\log(|\lambda_i(t)|).
\end{equation}
\begin{figure}[!h]
    \centering
    \subfloat[]{\includegraphics[width=0.34\linewidth]{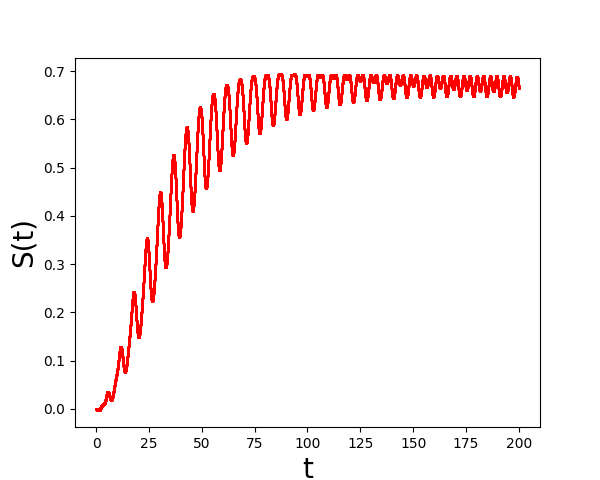}}
    \subfloat[]{\includegraphics[width=0.34\linewidth]{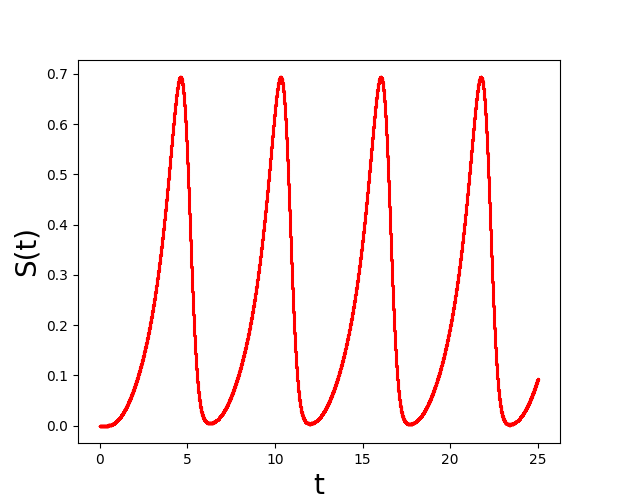}} \subfloat[]{\includegraphics[width=0.35\linewidth]{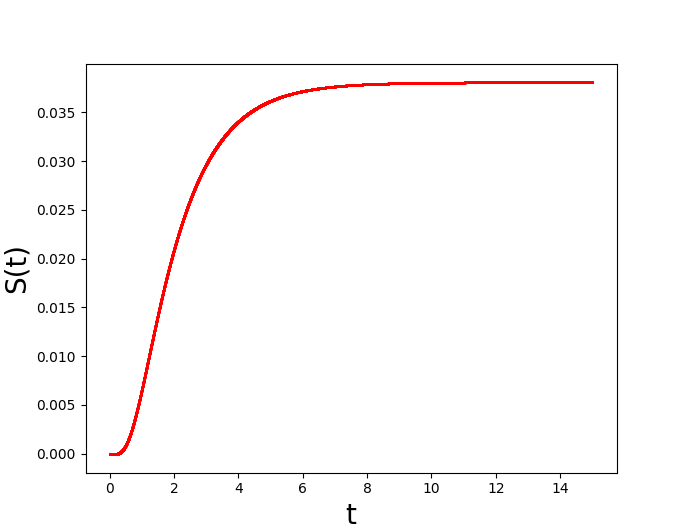}}  
    \caption{Figure shows the von Neumann entropy with respect to time (ns) for Rabi frequencies $2\omega_1 = 2\omega_2 = 1$, with both the qubits initially prepared in the $\ket{0}$ state. The plots are shown for the measurement frequencies $\alpha_1=\alpha_2=$ (a) 0.1, (b) 1.5, and (c) 3.}
    \label{fig:Entropy}
\end{figure}
For the case discussed above, we have plotted entanglement entropy with time in Fig. \ref{fig:Entropy} for the different values of the measurement frequencies $\alpha$ with the Rabi frequencies of qubits considered as $2\omega_1 = 2\omega_2 = 1$ GHz. The initial states of both the qubits are considered as $\ket{0}$. We observe that when the measurement frequency is smaller than the Rabi frequency, $\alpha = 0.1$, $S(t)$ oscillates and after a long time ($75$ ns), it approaches the maximum value $\ln \,2$ leading to the maximally entangled state. As the measurement frequency is slightly higher than the Rabi frequency, $\alpha = 1.5$, the entanglement entropy oscillates between $0$ and $\ln \,2$. In this case, the entanglement entropy reaches from one minima (maxima) to the next minima (maxima) after each $5.77$ ns. When the measurement frequency is increased further, $\alpha = 3$, the entanglement entropy saturates at $0.0377$ after $5$ ns (corresponding to the state $\ket{\psi} =  0.850147 i \ket{00} + 0.369797 (\ket{01} + \ket{10}) - 0.612372 i \ket{11}$).

Generally, the density matrix of the two-qubit system can be expressed at any given time as Eq. \ref{rhot}. First, we find the reduced matrix ($\rho_1$) using a partial trace which is the general single qubit density matrix in terms of Bloch sphere coordinates: $\rho_1 = (1/2)(\mathbb I + x_1 \sigma_x + y_1 \sigma_y + z_1 \sigma_z)$. We can find the analytic expression of entanglement entropy as:
\begin{alignat}{1}
S& = -{\rm tr} (\rho_1 \ln{\rho_1})\nonumber\\
&= \ln 2  - r_1 \tanh ^{-1} r_1 - \ln \sqrt{1-r_1^2},
\end{alignat}
where $r_1 = \sqrt{x_1^2+y_1^2+z_1^2}$. We know the composite system is separable if the reduced matrices are pure, consequently making the entropy zero, while a maximally entangled system leads to a maximally mixed reduced matrix, which would make the entropy $\ln \,2$. Analyzing the values of $r_1$ at these limits shows us that $r_1 = 1$ represents a separable state, and $r_1 = 0$ represents a maximally entangled state. Thus, we can get an idea of the entanglement of the system merely by calculating the lengths of the vectors  $\Bar{r}_1$ and $\Bar{r}_2$ without concerning ourselves with the entanglement variables $e_{ij}$.

\section{Achieving a target state with the quantum Zeno effect}\label{target state}
Measurement can be used to get the desired target state of a quantum system even when the system's initial state is unknown. Instead of performing the projective measurement on a quantum system (here, a qubit), we couple the system to a detector for a short duration, $\delta t$. Then switch off the interaction between the system and the detector and perform an instantaneous projective measurement of the detector. The measurement outcome tells the state of the system and measurement back action affects the system state due to entanglement. Here, we use the interaction strength and measurement back action to get the desired target state of the given quantum system. 

We consider a system of qubits interacting with an ancilla where the ancilla is being measured after each time interval $dt$ in the continuous measurement regime $dt \to 0$. After each measurement, the ancilla is reset to its initial state $\ket{0}_d$.  At any time, the state of the qubit can be represented in terms of an orthogonal basis. The Hamiltonian of this system with an ancilla can be written as:
\begin{equation}
    H = H_s + H_a + H_{\rm int}, 
\end{equation}
where $H_s$ represents the Hamiltonian of the qubit system, $H_a$ represents the Hamiltonian of the ancilla, and $H_{\rm int} = J\hbar(S\otimes A^\dagger + S^\dagger \otimes A)$ is the interaction Hamiltonian between the system and the ancilla with $ S(A)$ is an operator which acts on the qubit system (ancilla) and $J = \sqrt{\alpha/dt}$. The ancilla is being reset after each measurement. So, at any time $t$, the density matrix of the system and ancilla just after the measurement can be written as $\rho(t) = \rho_s(t)\otimes \rho_d$, where $\rho_s(t)$ is the density matrix of the qubit system and $\rho_d = \ket{0}\bra{0}$ is the density matrix of the ancilla. The density  matrix of the qubit after each measurement is given by:
\begin{equation}
    \rho_s(t+dt)=\frac{M^r U\rho_s(t)U^\dagger M^{r^\dagger}}{Tr[M^r U\rho_s(t)U^\dagger M^{r^\dagger}]},
\end{equation}
where  $r$ is the measurement outcome, $U=e^{-i H_s dt/\hbar}$ and $M^r=_d\bra{r}e^{-i H_{int}dt/\hbar}\ket{0}_d$ is the Kraus operator, $\sum_{r}M^{r^\dagger}M^r={\mathbb I}$. 

In the scaling limit for continuous measurement, $\delta t\to 0$, $J^2\delta t\to\alpha=\rm{constant}$, the evolution of the system in the presence of interaction with an ancilla is  given by the following equation:
\begin{equation}
    \frac{d\rho_s}{dt} =-i [H_s, \rho_s(t)] + \alpha  \left[S \rho_s (t) S^\dagger - \frac{1}{2} \{S^\dagger S, \rho_s(t) \} \right].
    \label{Lindblad}
\end{equation}
In the case of qubits interacting with an ancilla, if the initial state of the ancilla is not an eigenstate of the measurement operator $A$, then the above equation gives
\begin{equation}
    \frac{d\rho_s}{dt} =-i [H_s, \rho_s] - \frac{\alpha}{2} \{S^\dagger S, \rho_s\},
    \label{Lindblad2}
\end{equation}
It has been shown that when the measurement frequency is higher than the frequency of the qubit system, the qubit state freezes; exhibiting the quantum Zeno effect \cite{krj}. 

We want to use the quantum Zeno effect to realize the desired state. For which we have to go through these steps:\newline
(1) define the desired state $\rho_s(t)$,\newline
(2) find the required interaction operator $S$ such that $d\rho_s/dt = 0$,\newline
(3) with the interaction operator, design the interaction Hamiltonian $H_{int}$ and measurement frequency.

In the following, after recollecting weak measurement of a single qubit, we proceed to discuss the case of two qubits. 
\subsection{Weak measurement of a qubit}
First, we consider the weak continuous repeated measurement of single qubit that is performing coherent oscillations between states $\ket{0}$ and $\ket{1}$ with frequency $\omega$ and interacting with an ancilla. After each measurement, the ancilla is reset to its initial state $\ket{0}_d$ which is the eigenstate of the Pauli matrix $\sigma_{z}$.  The Hamiltonian of the qubit is $H_s = \hbar \omega \sigma_x/2$. At any time, we can write the density matrix of a qubit in terms of Bloch sphere coordinates $(x, y, z)$ and the Pauli matrices as $\rho_s(t)=(1/2)({\mathbb I} + x \sigma_x + y \sigma_y + z \sigma_z)$. Between the consecutive measurements, qubit and ancilla interact and evolve together. We consider the interaction as a Hermitian operator, written as:
\begin{alignat}{1} \label{Int}
 S & = \begin{pmatrix}
  a & b + i d\\
  b - i d & c
  \end{pmatrix}.
\end{alignat}
We find the updated equations of the Bloch sphere coordinates using \eqref{Lindblad2} as:
\begin{alignat}{1}
    \dot x & = \frac{(a + c)\alpha}{2}\left[2 b (x^2 - 1) - x \{ 2 d y + (-a + c) z\}\right]\nonumber\\
    \dot y & = \frac{(a + c) \alpha}{2} \left[ 2d + 2b x y -y\{2 d y + (-a+c) z\}\right] - \omega z\nonumber\\
    \dot z & =  \frac{(a + c) \alpha}{2} \left[ -a + c + 2 b x z -2 d y z + (a-c) z^2 \right] + \omega y.
\end{alignat}
The stationary solution of the above equations, $\dot x = 0$, $\dot y = 0$, and $\dot z = 0$ gives the coupled expressions for Bloch sphere coordinates of the target state in terms of the interaction parameters and the measurement frequency. For a particular choice of interaction parameter, $a=0$, $b=0$, $c=1$ and $d=0$, the stationary solutions give $x=0$, $y = -1/\lambda$, and $z = \sqrt{\lambda^2 - 1}/\lambda$ which is consistent with the result obtained in the \cite{krj}.

Let us consider the dynamics of a qubit in $y-z$ plane, i.e., $\phi = \pi/2$ such that $x = 0$,  $y=\sin{\theta}$ and $z=\cos{\theta}$, that gives 
\begin{alignat}{1}     
 \rho_s (t) & = \frac{1}{2}\begin{pmatrix}
  1 + \cos{\theta} & - i \sin{\theta}\\
  i \sin{\theta}  & 1-\cos{\theta}
  \end{pmatrix}.
\end{alignat}
The evolution equation of the  density matrix using \eqref{Int} and  \eqref{Lindblad2} turns out to be:
\begin{alignat}{1}
\frac{1}{2}\begin{pmatrix}
  -\sin {\theta}\hspace{0.5mm} \dot {\theta} & - i \cos {\theta}\hspace{0.5mm} \dot {\theta}\\
  i \cos {\theta}\hspace{0.5mm} \dot {\theta} & \sin {\theta}\hspace{0.5mm} \dot {\theta}\nonumber\\
\end{pmatrix}
=
\frac{1}{4}\begin{pmatrix}
\beta & \gamma + i \delta \\
\gamma - i \delta  & -\beta,
\end{pmatrix}
\end{alignat}
where $\beta= \sin{\theta} [2\omega -(a+c) \alpha \{2 d \cos { \theta} + (a - c) \sin{\theta}\}]$,  $\gamma = -2 (a + c)b$ and $\delta = \cos{\theta} [2\omega -(a+c) \alpha \{2 d \cos { \theta} + (a - c) \sin{\theta}\}]$. Comparing the two sides, we require:
\begin{alignat}{1}
    \text{either} \quad b& = 0\quad\text{or}\quad a = -c\quad \text{which give} \nonumber\\
    \dot{\theta}& = -\frac{1}{2} \left[2\omega -(a+c) \alpha \{2 d \cos { \theta} + (a - c) \sin{\theta}\}\right]\quad \text{and}\quad \dot{\theta}=-\omega.
\end{alignat}
To achieve a target state, i.e., $\dot {\theta} = 0$, $b=0$ is required. And we get the target value of $\theta$ in terms of $a, c, d$ and $\lambda$:  
\begin{equation}
    \theta = \sin ^{-1} \bigg[\frac{1}{\lambda (a+c) \sqrt{4 d^2 + (a-c)^2}}\bigg]-\xi,
\end{equation}
where $\lambda = \alpha/ (2\omega) $ is the ratio of the measurement frequency to the Rabi frequency of the qubit and $\tan{\xi} = 2d/(a-c)$. By fixing the value of desired $\theta$, we can design the interaction operator by choosing the values of $a, c$ and $\lambda$. 

We can choose $a=0$, $c=1$, and $d=0$, i.e., the ancilla is monitoring the state $\ket{1}$ of the qubit, it gives an expression of the critical value of $\theta$  as $\theta = \sin^{-1}\left(1/\lambda \right), -\sin^{-1}\left(1/\lambda \right)$ as discussed in \cite{krj}. We can get the desired frozen state with these values by changing the value of $\lambda > 1 $. 
\subsection{System of two qubits}
Consider the case of continuous repeated weak measurements of a two-qubit system interacting with an ancilla. The Hamiltonian of this system with an ancilla can be written as:
\begin{equation}
    H = H_s + H_a + H_{int}, 
\end{equation}
where $H_s = (\sigma_z \otimes {\mathbb I} + {\mathbb I}\otimes \sigma_z)$ represents the Hamiltonian of the two-qubit system, $H_a$ represents the Hamiltonian of the ancilla, and $H_{int} = J\hbar\sum_{j=1}^{4}(S_j A^\dagger + L_j^\dagger A)$ is the interaction between the system and the ancilla with $A$ as an operator which acts on the two-qubit system (ancilla). We write the orthogonal basis of the two-qubit system state as  $ \{\ket{B_1}, \ket{B_2}, \ket{B_3}, \ket{B_4}\}$ with $\ket{B_1}= \ket{00}$ ($\ket{B_2}= \ket{01}$, $\ket{B_3}= \ket{10}$, and $\ket{B_4}= \ket{11}$). 

Consider that the target state $\rho_s(t)$ given by $\ket{\psi_S(t)} = \ket{00}$ or $\ket{11}$ with $H_s=(\sigma_z \otimes {\mathbb I} + {\mathbb I}\otimes \sigma_z)$ for which $[H_s, \rho_s]=0$. We have to design the interaction Hamiltonian to achieve this target state. First, we consider the Lindblad  jump operator (independent of the choice of measurement operator) as:
\begin{alignat}{1}
    L_j& = (1-\delta_{1j}-\delta_{4j})(\ket{B_1}\bra{B_j} + \ket{B_4}\bra{B_j})\quad \text{which is nonzero only for} \quad \text{j =2, 3},\nonumber\\
   L_j^\dagger L_j& = (\ket{B_j}\bra{B_1} + \ket{B_j}\bra{B_4})(\ket{B_1}\bra{B_j} + \ket{B_4}\bra{B_j})\nonumber\\ 
   & = 2\ket{B_j}\bra{B_j}.
\end{alignat}
 The substitution of this Lindblad jump operator with the required condition that $d\rho_s/dt = 0$ for the target state in \eqref{Lindblad} gives:
 \begin{alignat}{1}
     \sum_{j=2,3}\left[ (\ket{B_1}\bra{B_j} + \ket{B_4}\bra{B_j}) \rho_s(t)(\ket{B_j}\bra{B_1} + \ket{B_j}\bra{B_4}) - \{\ket{B_j}\bra{B_j},\rho_s(t)\}\right]=0.
 \end{alignat}
The above equation is satisfied for:
\begin{alignat}{1}
    \rho_s(t)& = \ket{00}\bra{00}, \quad \text{and} \quad\ket{11}\bra{11}
\end{alignat}
 which shows that the frozen state would be either $\ket{00}$ or $\ket{11}$ with equal probability. The required interaction Hamiltonian is:
\begin{equation}
    H_{int} = J\sum_{j=2,3} \{(\ket{B_1}\bra{B_j} + \ket{B_4}\bra{B_j})\otimes \sigma_d^\dagger + h.c \}.
\end{equation}

\section{Concluding remarks}
Quantum Zeno Effect  has been shown to provide an analogue of Dehmelt-shelving, thereby opening the possibility to control quantum jumps. This has been shown for a qubit weakly measured by an ancilla by repeated measurements \cite{krj}. Here we have shown how to employ QZE to shelve the state of a two-qubit system by making them interact with an ancilla using weak measurement. With control of two-qubit system, we can correct errors occurring in two-qubit gates, hence it is quite  significant. Also, this work opens the door to two-qubit two-ancilla system where some preliminary results have been found \cite{garima_2q2a}.

The freezing of the state is realized after a nontrivial evolution of entanglement of the qubits. This shows in the entanglement entropy calculated by employing one-qubit reduced density matrix shows saturation to maximal mixing; however, the approach to this is oscillatory or monotonic, depending on the ratio of measurement frequency and Rabi frequency. 

Moreover, a quantum system can be brought to a desired state by manipulating the interaction strength and measurement back action. This gives us how to engineer the interaction to realize a state. We believe that this could play an important role in helping the circuit engineering, and in developing encoding protocols.

\newpage
\appendix
\section*{Appendix}
The Hamiltonian that includes the effect of measurement on the two-qubit system interacting with an ancilla can be written as given in Eq.\ref{eq:ham}. We can find the updated equations of the generalized momenta using Hamilton's equation:
\begin{align}
\dot{p}_{x1}&=\frac{1}{2} \left(z_1 p_{x1} \alpha_1+((z_1+z_2-e_{33}) p_{x1}-p e_{13}) \sqrt{\alpha_1 \alpha_2} + (z_2 p_{x1}-p_{e13}) \alpha_2\right)\nonumber\\
\dot{p}_{y1}&=\frac{1}{2} \left(z_1 p_{y1} \alpha_1+((z_1+z_2-e_{33}) p_{y1}-p_{e23}) \sqrt{\alpha
1} \sqrt{\alpha_2}+(z_2 p_{y1} - p_{e23}) \alpha_2 - 4 p_{z_1} \omega_1\right)\nonumber\\
\dot{p}_{z1}&=\frac{1}{2} \Biggl[(-1+x_1 p_{x1} + y_1 p_{y1} + 2 z_1 p_{z1} + x_2 p_{x2} + y_2 p_{y2} + z_2 p_{z2} + e_{11} p_{e11} + e_{12} p_{e12}\nonumber\\
& + e_{13} p_{e13} + e_{21} p_{e21} + e_{22} p_{e22} + e_{23} p_{e23} + e_{31} p_{e31} + e_{32} p_{e32} + e_{33} p_{e33}) \alpha_1\nonumber\\
  &+(-1+x_1 p_{x1} + y_1 p_{y1} + 2 z_1 p_{z1} - e_{33} p_{z1} + x_2 p_{x2} +  y_2 p_{y2} + p_{z2} + z_2 (p_{z1} + p_{z2}) + e_{11} p_{e11}\nonumber\\
  & + e_{12} p_{e12} + e_{13} p_{e13} + e_{21} p_{e21} + e_{22}p_{e22} + e_{23} p_{e23} + e_{31} p_{e31} + e_{32} p_{e32}\nonumber\\
  &+(-1 + e_{33}) p_{e33})\sqrt{\alpha_1 \alpha_2}  + z_2 p_{z1} \alpha_2 - p_{e33} \alpha_2 + 4 p_{y1} \omega_1\Biggr]\nonumber\\
\dot{p}_{x2}& = \frac{1}{2} \left((z_1 p_{x2} - p_{e31}) \alpha_1 + ((z_1 + z_2 - e_{33}) p_{x2} - p_{e31})\sqrt{\alpha_1 \alpha_2} + z_2 p_{x2} \alpha_2\right)\nonumber\\
\dot{p}_{y2}&=\frac{1}{2} \left((z_1 p_{y2} - p_{e32}) \alpha_1 + ((z_1 + z_2 - e_{33}) p_{y2} - p_{e32})
\sqrt{\alpha_1 \alpha_2} + z_2 p_{y2} \alpha_2 - 4 p_{z2} \omega_2\right)\nonumber\\
\dot{p}_{z2}&=\frac{1}{2} \Biggl[\sqrt{\alpha_1 \alpha_2} (-1+x_1 p_{x1} +y_1 p_{y1} +p_{z1} +p_{x2}
p_{x2} +y_2 p_{y2} +2 z_2 p_{z2}-e_{33} p_{z2} +e_{11} p_{e11}\nonumber\\
&+e_{12} p_{e12} +e_{13} p_{e13}+e_{21} p_{e21} +e_{22} p_{e22}+e_{23} p_{e23} +e_{31} p_{e31}+e_{32} p_{e32})\nonumber\\
&+\alpha_2(-1+x_1 p_{x1} +y_1 p_{y1} +p_{x2} p_{x2} +y_2 p_{y2} +2 z_2 p_{z2} +e_{11} p_{e11} +e_{12} p_{e12} +e_{13} p_{e13}\nonumber\\
&+e_{21} p_{e21} +e_{22} p_{e22} +e_{23} p_{e23} +e_{31} p_{e31} +e_{32} p_{e32}) +p_{e33}(-\alpha_1+(-1+e_{33}) \sqrt{\alpha_1 \alpha_2} \nonumber\\
&+e_{33} \alpha_2)+z_1 \left(p_{z2} \alpha_1+(p_{z1}+p_{z2}) \sqrt{\alpha_1} \sqrt{\alpha
2}+p_{z1} \alpha_2\right)+4 p_{y2} \omega_2\Biggr]\nonumber\\
\dot{p}_{e11}&=\frac{1}{2} \left(z_1 p_{e11} \alpha_1 + ((z_1+z_2-e_{33}) p_{e11}-p_{e22}) \sqrt{\alpha_1 \alpha_2} + z_2 p_{e11} \alpha_2\right)\nonumber\\
\dot{p}_{e12}&=\frac{1}{2} \left(z_1 p_{e12} \alpha_1+((z_1+z_2-e_{33}) p_{e12}+p_{e21}) \sqrt{\alpha_1 \alpha_2} + z_2 p_{e12} \alpha_2-4 p_{e13} \omega_2\right)\nonumber\\
\dot{p}_{e13}&=\frac{1}{2} \left(z_1 p_{e13} \alpha_1+(-p_{x1}+(z_1+z_2-e_{33}) p_{e13}) \sqrt{\alpha_1 \alpha_2} - p_{x1} \alpha_2+z_2 p_{e13} \alpha_2+4 p_{e12} \omega_2\right)\nonumber\\
\dot{p}_{e21}&=\frac{1}{2} \left(z_1 p_{e21} \alpha_1+(p_{e12}+(z_1+z_2-e_{33}) p_{e21}) \sqrt{\alpha_1 \alpha_2} + z_2 p_{e21} \alpha_2-4 p_{e31} \omega_1\right)\nonumber\\
\dot{p}_{e22}&=\frac{1}{2} \left(z_1 p_{e22} \alpha_1+(-p_{e11}+(z_1+z_2-e_{33}) p_{e22})\sqrt{\alpha_1 \alpha_2} + z_2 p_{e22} \alpha_2-4 (p_{e32} \omega_1+p_{e23} \omega_2)\right)\nonumber
\end{align}
\begin{align}
\dot{p}_{e23}&=\frac{1}{2}(z_1 p_{e23} \alpha_1+(-p_{y1}+(z_1+z_2-e_{33}) p_{e23})\sqrt{\alpha_1 \alpha_2} - p_{y1} \alpha_2+z_2 p_{e23} \alpha_2-4 p_{e33} \omega_1\nonumber\\
  &+4 p_{e22}\omega_2)\nonumber\\
\dot{p}_{e31}&=\frac{1}{2} \left(-p_{x2} \alpha_1+z_1 p_{e31} \alpha_1+(-p_{x2}+(z_1+z_2-e_{33})
p_{e31}) \sqrt{\alpha_1 \alpha_2} +z_2 p_{e31} \alpha_2+4 p_{e21} \omega_1\right)\nonumber\\
\dot{p}_{e32}&=\frac{1}{2} (-p_{y2} \alpha_1+z_1 p_{e32} \alpha_1+(-p_{y2}+(z_1+z_2-e_{33})
		p_{e32}) \sqrt{\alpha_1 \alpha_2} +z_2 p_{e32} \alpha_2+4 p_{e22} \omega_1\nonumber\\
  &-4 p_{e33}\omega_2)\nonumber\\
\dot{p}_{e33}&=\frac{1}{2} \bigg[(-p_{z2} \alpha_1+z_1 p_{e33} \alpha_1+\sqrt{\alpha_1 \alpha_2} (1-x_1 p_{x1} -y_1 p_{y1} -p_{z1}
-p_{x2} p_{x2} -y_2 p_{y2}\nonumber\\
&-(1+z_2) p_{z2} -e_{11} p_{e11} -e_{12} p_{e12} -e_{13} p_{e13} -e_{21} p_{e21} -e_{22} p_{e22} -e_{23} p_{e23} -e_{31} p_{e31}\nonumber\\
&-e_{32} p_{e32} +z_2 p_{e33} -2 e_{33} p_{e33} +z_1 (-p_{z1}+p_{e33}))-p_{z1} \alpha_2+z_2 p_{e33} \alpha_2+4 p_{e23} \omega_1+4 p_{e32}
\omega_2 \bigg]
\end{align}

\end{document}